\documentclass[aps,pra,twocolumn,superscriptaddress,showpacs,10pt]{revtex4-1}
\usepackage{amssymb}
\usepackage{graphicx}
\usepackage{amsmath}
\usepackage{verbatim,dsfont,color}

\begin{document}

\title{Diabolical points in multi-scatterer optomechanical systems}

\author{Stefano Chesi}
\email[]{Correspondence to: stefano.chesi@csrc.ac.cn}
\affiliation{Beijing Computational Science Research Center, Beijing 100084, China}

\author{Ying-Dan Wang}
\affiliation{State Key Laboratory of Theoretical Physics, Institute of Theoretical Physics,
Chinese Academy of Sciences, P.O. Box 2735, Beijing 100190, China}

\author{Jason Twamley}
\affiliation{Centre for Engineered Quantum Systems, Department of Physics and Astronomy,
Macquarie University, NSW 2109, Australia}

\date{\today}

\begin{abstract}

Diabolical points, which originate from parameter-dependent accidental degeneracies of a system's energy levels, have played a fundamental role in the discovery of the Berry phase as well as in photonics (conical refraction), in chemical dynamics, and more recently in novel materials such as graphene, whose electronic band structure possess Dirac points. Here we discuss diabolical points in an optomechanical system formed by multiple scatterers in an optical cavity with periodic boundary conditions. Such configuration is close to experimental setups using micro-toroidal rings with indentations or near-field scatterers. We find that the optomechanical coupling is no longer an analytic function near the diabolical point and demonstrate the topological phase arising through the mechanical motion. Similar to a Fabry-Perot resonator, the optomechanical coupling can grow with the number of scatterers. We also introduce a minimal quantum model of a diabolical point, which establishes a connection to the motion of an arbitrary-spin particle in a 2D parabolic quantum dot with spin-orbit coupling.
\end{abstract}





\maketitle

\section*{Introduction} 

Accidental degeneracies are a generic phenomenon for quantum mechanical problems depending on at least two external parameters \cite{vonNeumann:1929ta,Teller:1937vi}. Such degeneracies can appear in sets of codimension-2 for $d$-parameterized Hamiltonians and for $d=2$ these sets have measure zero and are isolated points. The intersecting energy sufaces at these points form a cone and such conical intersections are termed {\em diabolical points} (DPs), after the corresponding shape of the juggler's diabolo top. Initially discovered within optics by Hamilton  in the conical refraction of light \cite{Hamilton:1831vv}, DPs  were made prominent by Berry \cite{Berry1984, Berry1984b}, who showed that a system acrues a phase when it evolves adiabatically through a closed path in  parameter space enclosing the DP: the Berry phase, or more precisely, a topological Berry phase \cite{EriccsonPhD2002}. 

Due to such topological phase and their peculiar energy dispersion, DPs have numerous important consequences, for example in molecular reactions \cite{Yarkony1996}, conical refraction in crystal optics \cite{Berry:2004gs, Berry:2007iw}, molecular spectra \cite{Mead:1979id,Cederbaum:2003fk}, neutral atoms in optical lattices \cite{Wuster2011,Zhang:2011ef,Tarruell:2013db}, and honeycomb photonic systems  \cite{Peleg:2007ep, BahatTreidel:2008uo, Rechtsman:2013cz}. When entire electronic bands possess DPs, such as discovered in graphene and topological insulators, in their vicinity the electrons behave as massless fermions obeying the Dirac equation \cite{Novoselov:2005es, Zhang:2005gp}, and the diabolical points are known as Dirac points. This discovery has generated enormous interest and there are many works generalizing the graphene-type model to other physical systems and exploring the associated topological properties \cite{2007NatMa...6..183G, Rechtsman:2014fe}. For example, the $\pi$ topological phase of such Dirac fermions affects the quantum interference corrections to the conductivity, as found in experiments on topological insulators (see Ref.~\cite{Hai-hou2104_refs}, and references therein). 

We study here diabolical points in the coordinate space of cooperative many-body optomechanical systems. The specific realization we focus on is a one-dimensional ring cavity with an embedded Distributed Bragg Reflector (DBR)  formed by moveable individual reflectors. This setup can be treated analytically, and is directly applicable to an array of membranes in a ring cavity, see Fig.~\ref{fig:schematics}(a), or a toroidal cavity with indentations~\cite{Arbabi:2011ec}, see Fig.~\ref{fig:schematics}(b).  More generally, the DBR could be replaced by any type of composite scattering elements (e.g., collections of near-field oscillators \cite{Anetsberger:2009sp}, nanoprobes \cite{Zhu2010}, or atomic clouds \cite{Botter:2013ki}). Mode interactions in Fabry-Perot cavities could give rise to DPs by taking into account three-dimensional features, e.g., the tilt-angle of the scattering elements \cite{Sankey2009}. However, within the 1D model considered here, periodic boundary conditions are a necessary feature to observe DPs.

Besides their spectroscopic characterization, DPs are most interesting for their consequences on the system's dynamics. In particular, we show that a topological Berry phase can be realized by different types of parameterized cyclic motion of the scatterers. While a geometric phase generally takes arbitrary continuous values, depending on the detailed local structure of the closed adiabatic path (see Ref. \onlinecite{Khosla2013}  for an example in optomechanics), in the special case of a DP the associated Berry phase is {\em only} a function of the winding number around the DP. For this reason this type of Berry phase is non-local, i.e., any homologically equivalent path encircling the DP acquires identical discrete phases: it is a more robust phenomenon denoted as a topological phase and bears a strict relation with the Aharonov-Bohm or Aharonov-Casher effect (see, e.g., Ref.~\cite{EriccsonPhD2002}). The spectrum and topological phase allow to demonstrate the presence of DPs in experiments and to distinguish them from exceptional points, a different type of singularity recently discussed in optomechanical systems \cite{Zhu2010,Wiersig2011,Wiersig2014}.

For mechanical motion at the DP, the transduction of photons between optical branches becomes possible, controlled by Landau-Zener physics \cite{Teller:1937vi}. Noticeably, near a DP the standard perturbative derivation of the optomechanical interaction - where the photon number couples to the mechanical position (linear or quadratic) - breaks down and one must resort to a more fundamental description of the optomechanical interaction. For this reason we consider an explicit optomechanical model which allows us to address in more detail the quantum behavior around a DP. Such model of a DP also establishes an analogy to parabolic quantum dots in semiconductors with a general form of spin-orbit coupling (Rashba plus Dresselhaus \cite{Bychkov1984,Dresselhaus1955}). In the optomechanical system, however, the `electron' spin is allowed to have an arbitrary value $J\ge 1/2$, determined by the total number of photons. 
 

\section*{Results} \label{sec:system}

\subsection*{Occurrence of diabolical points}

\begin{figure}
\begin{center}
\includegraphics[width=0.48\textwidth]{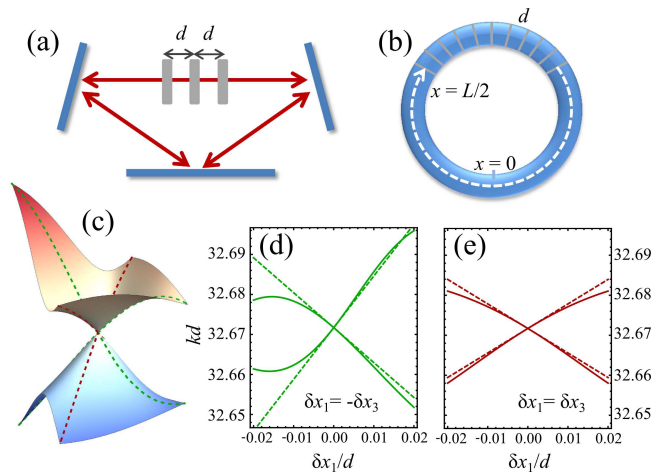}
\end{center}
\caption{\label{schematics} (a) Schematics of a ring cavity with embedded DBR. (b) Alternative setup, based on a ring resonator (see, e.g., \cite{Arbabi2011}). (c) Spectrum of a cavity with $3$ scatterers (middle one fixed) as function of the displacements of the two side scatterers ($\delta x_1$ and $\delta x_3$). Two optical branches show a conical intersection (diabolical point) in the vicinity of the equilibrium point ($\delta x_1=\delta x_3=0$). Panels (d)-(e) are cross-sections along the dashed lines of panel (c), i.e., for $\delta x_1=-\delta x_3$ (d) and  $\delta x_1=\delta x_3$ (e). The solid lines in (d) and (e) are obtained numerically while the dashed lines are analytical results from Eq.~(\ref{dk_general}). The parameters assumed in plots (c)-(e) are $n=1$, $\zeta=-0.2$, $\phi=0$, and $L/d=19.17$.  }
\label{fig:schematics}
\end{figure}

We consider $N$ identical movable scattering elements embedded in a ring cavity of refractive index $n$, as shown in Fig. \ref{schematics}. If each scatterer is assumed to be symmetric, it can be characterized by a real polarizability $\zeta \equiv r/{i t}$ and a phase factor $e^{i \phi}\equiv(1-i\zeta)t$ (where $r$ is the reflectivity and $t$ the transmittivity). In the equilibrium position, the spacing between the scatterers is $d$ and the cavity length excluding the DBR is $L$. We fix the origin of coordinates in the center of the $L$ section, such that the first scatterer is at $x=L/2$, see Fig.~\ref{fig:schematics}(b). Under these assumptions, using the transfer matrix for a single scatterer:
\begin{equation} \label{T_scatterer}
{\bf M}_{\rm s}=\left( 
\begin{array}{cc}
\left( 1+i\zeta \right) e^{i\phi } & i\zeta  \\ 
-i\zeta  & \left( 1-i\zeta \right) e^{-i\phi }%
\end{array}%
\right),
\end{equation}%
and for free propagation on a distance $\Delta x$:
\begin{equation}  \label{T_vacuum}
{\bf M}_{\rm p}=\left( 
\begin{array}{cc}
e^{ink\Delta x} & 0 \\ 
0 & e^{-ink\Delta x}%
\end{array}%
\right),
\end{equation}
the following total transfer matrix of the cavity is obtained:
\begin{equation}
\mathbf{M}=\left( 
\begin{array}{cc}
\left( 1+i\chi \right) e^{i\varphi } & i\chi \\ 
-i\chi & \left( 1-i\chi \right) e^{-i\varphi }%
\end{array}%
\right),  \label{Mt}
\end{equation}%
with the effective polarizability $\chi =\zeta U_{N-1}\left( a\right) $, where $a=\cos
\left(nkd+\phi \right) -\zeta \sin \left( nkd+\phi \right) $, and $%
U_m\left( x\right) $ is the $m$-th Chebyshev polynomial of
the second kind, which satisfies $U_{N-1}\left( \cos \lambda \right) =\sin N\lambda /\sin
\lambda $. The effective phase factor is
\begin{equation} 
e^{i\varphi }=\frac{e^{i(nkL+\phi)}[1-i\zeta U_{N-1}\left( a\right)] }{\left( 1-i\zeta \right)
U_{N-1}\left( a\right) -e^{i\left( nkd+\phi \right) }U_{N-2}\left( a\right) }.
\end{equation}

In order to describe the motion of the mechanical elements around the equilibrium position (while keeping the total length of the cavity $L_{\rm tot}=L+(N-1)d$ unchanged) we consider the cavity transfer matrix $\mathbf{M}(k,\delta \vec{x})$ for general coordinates of the scatterers, where  $\delta\vec{x}$ is the $N$-dimensional displacement vector of the scatterers from their equilibrium position. The transfer matrix $\mathbf{M}(k,\delta \vec{x})$ is derived by using Eqs.~(\ref{T_scatterer}) and (\ref{T_vacuum}) and the wavevectors $k$ of the optical modes satisfy the periodic boundary conditions:
\begin{equation}
\mathbf{M}\mathbf{v}=\mathbf{v}, \label{bc}
\end{equation}%
for at least one  amplitude vector ${\bf v}^T=(A_L,A_R)$. 

We can now characterize the diabolical points of the periodic cavity, which occur at certain values of $k$ corresponding to accidental degeneracies of the optical modes. The clockwise/counterclockwise waves, ${\bf v}^T=(1,0)/(0,1)$ in Eq.~(\ref{bc}), are both eigenmodes of the system at a diabolical point, implying that $\mathbf{M}={\bf 1}$ for this $k$. Clearly, for the type of one-dimensional model considered here, a degeneracy of the spectrum is impossible in a Fabry-Perot setup since the left/right-propogating waves are necessarily coupled by the end mirrors and anticrossing in the energy spectrum is unavoidable. 

We consider first the equilibrium configuration of equidistant scatterers, for which the cavity transfer matrix is given by Eq.~(\ref{Mt}). Then, the condition $\mathbf{M}={\bf 1}$ requires the DBR to be transparent ($\chi =0$) and that the total phase satisfies $e^{i\varphi}=1$. These two conditions give:
\begin{equation}
ndk_{0} =2p\pi -\phi - i \ln\left(\frac{\cos \theta _{m}+ i \sigma \zeta g_s}{1+i\zeta }\right) \equiv \nu_{m,p,\sigma},  
\label{diapcond1} 
\end{equation}%
and
\begin{equation}
L_{\rm tot} =Nd+\frac{\pi }{k_{0}n}\left( 2p'+m\right),  \label{diapcond}
\end{equation}
where  $\theta _{m}=m\pi /N$, $g_s^2=1+(\sin^2\theta_m)/\zeta^2$, $\sigma=\pm$, and $m$, $p$, $p'$ are integers.

\subsection*{Motion of scatterers and linearization}\label{sec:linearization}

If Eqs. (\ref{diapcond1}) and (\ref{diapcond}) are satisfied, the equilibrium position with equidistant scatterers coincides with the DP. To find the displacement-dependent energy spectrum around such a DP, one should solve  Eq.~(\ref{bc}) with the displacement-dependent transfer matrix $\mathbf{M}(k,\delta \vec{x})$. As an example, the two optical branches for a $N=3$ cavity with the middle scatterer fixed  ($\delta x_2=0$) are shown in Fig.~\ref{fig:schematics} in a small region around $\delta x_1=\delta x_3=0$. Here we have chosen parameters such that the equilibrium position is a diabolical point. 

When the scatterers are moved away from the diabolical point, either parametrically or dynamically, the two-fold degeneracy of the unperturbed wavevector  $k_0$ is lifted and the two optical modes have $k_{\pm }(\delta \vec{x})= k_0+\delta k_{\pm }(\delta \vec{x})$. We thus perform a linear expansion of the cavity transfer matrix:
\begin{equation}\label{dM_linearized}
\mathbf{M}(k,\delta\vec{x})\simeq {\bf 1}+\sum_{i=1}^N\frac{\partial \mathbf{M}}{%
\partial \delta x_{i}}\delta x_{i}+\frac{\partial \mathbf{M}}{\partial k}\delta
k ,
\end{equation}
which, together with Eq.~(\ref{bc}), immediately leads to the following equation:
\begin{equation}
\left( \frac{\partial \mathbf{M}}{\partial k}\right) ^{-1}\left( \sum_{i}%
\frac{\partial \mathbf{M}}{\partial x_{i}}\delta x_{i}\right) {\bf v}_\pm
=-\delta k_\pm{\bf v}_\pm.
\label{bc_linearized}
\end{equation}%
The explicit dependence of $\delta k_{\pm }(\delta \vec{x})$ can be found from Eq.~(\ref{bc_linearized}) after obtaining suitable expressions for $\frac{\partial \mathbf{M}}{\partial \delta x_{i}}$ and $\frac{\partial \mathbf{M}}{\partial k}$. The coordinate variation of  $\mathbf{M}(k,\delta \vec{x})$ can be computed following the method discussed in \cite{Xuereb2012}:
\begin{equation}
\frac{\partial \mathbf{M}}{\partial \delta x_{i}}=\left( 
\begin{array}{cc}
\alpha _{j}e^{ink_{0}L} & \beta _{j} \\ 
\beta _{j}^{\ast } & \alpha^*_{j}e^{-ink_{0}L}%
\end{array}%
\right),
\label{dMdx}
\end{equation}
where we have obtained, with $\theta^{(j)}_{m}=(2j-1)\theta_m$:
\begin{eqnarray}
&&\alpha _{j} =-2ik_{0}n\zeta ^{2}\frac{\sin \theta^{(j)}_{m}}{\sin \theta _{m}}e^{-ink_{0}L}  , \\
&&\beta _{j} = -2k_{0}n\zeta  \left( -1\right) ^{m}  \bigg( \cos \theta^{(j)}_{m}
 +i \sigma \zeta g_s \frac{\sin \theta^{(j)}_{m}}{\sin \theta _{m}}\bigg). \quad
\end{eqnarray}
The wavevector derivative is obtained as:
\begin{equation}
\frac{\partial \mathbf{M}}{\partial k}=\left( 
\begin{array}{cc}
i\left( \mu^{\prime }+\chi^{\prime }+nL\right) & i\chi^{\prime } \\ 
-i\chi^{\prime } & -i\left( \mu^{\prime }+\chi^{\prime
}+nL\right)%
\end{array}%
\right),\label{dMdk}
\end{equation}
where 
\begin{equation}
\chi^{\prime } \equiv \left. \frac{\partial \chi }{\partial k}%
\right\vert _{k=k_{0}}=\sigma \left( -1\right) ^{m}\frac{\zeta^2 Nnd}{\sin
^{2}\theta _{m}} g_s,
\end{equation}%
and $\mu=\varphi-\phi-nk_0L$ is the phase shift due to the unperturbed DBR, whose derivative is:
\begin{equation}
\mu^{\prime } \equiv \left. \frac{\partial \mu }{\partial k}%
\right\vert _{k=k_{0}}=\left( N-1\right) nd+\frac{\zeta^2 Nnd}{\sin ^{2}\theta
_{m}}\left( 1-\sigma \left( -1\right) ^{m} g_s\right).
\end{equation}%

Finally, we can plug these expressions into Eq.~(\ref{bc_linearized}) to obtain the linearized spectrum at the DP:
\begin{equation}
\delta k_\pm= \frac{\sqrt{2N}k_0\sin \theta _{m}}{g_c^2 L_{\rm tot}}
\left[\left(1-\eta\right)\delta x_s \pm \sqrt{\sum_{\mu=s,c} (g_\mu \delta x_\mu)^{2}}\right].
\label{dk_general}
\end{equation}
with $\eta=Nd/L_{\rm tot}$, $g_c^2=g_s^2-(1-\eta)^2$, and
\begin{equation}
\delta x_s=\delta \vec{x} \cdot \hat{s},\quad \delta x_c=\delta \vec{x}\cdot\hat{c},
\end{equation}
where $\hat{s}, \hat{c}$ are unit vectors with components $s_{j}=\sqrt{2/N}\sin [\left(2j-1\right) \theta _{m}]$ and $c_{j}=\sqrt{2/N}\cos [\left( 2j-1\right) \theta _{m}]$ ($j=1,2,\cdots,N$). The structure of Eq.~(\ref{dk_general}) shows that the degeneracy is lifted to linear order in the plane defined by $\hat{s}, \hat{c}$, also called \emph{branching plane} (or \emph{g-h plane}) \cite{Yarkony1996}, while the orthogonal $(N-2)$-subspace is the \emph{seam}, where the splitting is quadratic or zero (e.g., if all scatterers have the same displacement $\delta x_i={\rm const.}$, the degeneracy will not be lifted).

\begin{figure}
\begin{center}
\includegraphics[width=0.4\textwidth]{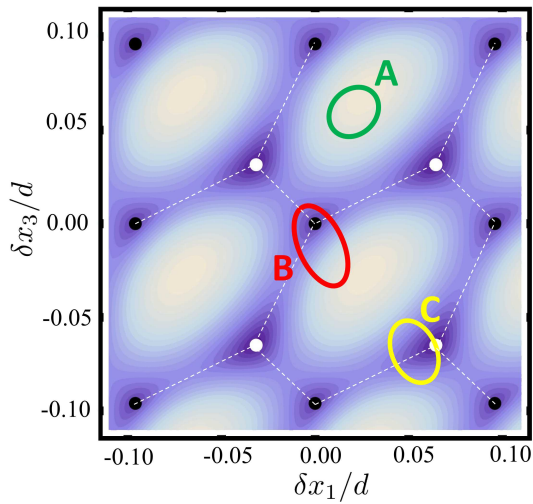}
\end{center}
\caption{Contour plot of $\delta k_+$ for the same setup of Fig. 1(c), but for a larger displacement range. The black (dark) and white (light) dots denotes two different families of DPs. Each family forms a square lattice with periodicity $\pi/k_0$, where the two values of $k_0$ are slightly different. For the black (dark) dots, $k_0=32.672$, obtained from Eq.~(\ref{diapcond}) with $p=5$,  $m=1$, and $\sigma = +$; for the white (light) dots, $k_0=32.669$, obtained from Eq.~(\ref{diapcond}) with $p=5$,  $m=2$, and $\sigma = +$. A different topological Berry phase $\Phi$ will be obtained by moving the scatterers along topologically different closed circuits: $\Phi=0$ for Loop A (green) and $\Phi=\pi$ for Loop B (red). Following the loop C (yellow) leads to the swap of $\pm$ photons.}
\label{fig:eigenmodes}
\end{figure}

In Fig.~\ref{fig:schematics}(d) and (e) we show that the linear approximation Eq.~(\ref{bc_linearized}) is accurate unless the displacements $\delta x_i$ are sufficiently large. In fact, by considering in Fig.~\ref{fig:eigenmodes} a larger range of displacements, we find a complex energy landscape for  $k_\pm$, with multiple DPs arranged in an approximately hexagonal pattern. More precisely, an eigenmode with unchanged frequency $k_0$ can be found by displacing any mirror by $\pi/k_0$. Therefore, the DPs of Fig.~\ref{fig:eigenmodes} form two square lattices (with slightly different periods). By changing the integer values $m$, $p$, $p'$, as well as $\sigma$, other energy sheets can be addressed. For example, when $\delta x_1=-\delta x_3 =\delta x$, Eqs.~(\ref{diapcond1}) and (\ref{diapcond}) allow us to find that DPs occur at:
\begin{equation}
\delta x=\frac{\nu_{m,p,\sigma}L_{\rm tot}}{\left( m+2p^{\prime
}\right) \pi +N\nu_{m,p,\sigma}}-d.
\end{equation}%

These conical intersections would be accessible, e.g., in experiments with ring cavities coupled to nanoprobes, where the ability to control accurately the scatterer positions has allowed to demonstrate degeneracies associated to exceptional points \cite{Zhu2010}. A DP can be distinguished from an exceptional point by the functional dependence of the mode splitting on displacements of the nanoprobes, as well as by the markedly different character of the optical modes: instead of approaching a single chiral mode \cite{Wiersig2011,Wiersig2014}, two degenerate counter-propagating modes coexist at the DP degeneracy.

\subsection*{Optomechanical couplings}

We now study in more detail the conical spectrum described by Eq.~(\ref{dk_general}). Probing the DP requires a two-dimensional motion: as seen from Eq.~(\ref{dk_general}), the change in the frequency of the optical modes has a non-analytic dependence on $\delta \vec x$, i.e., the conventional way to derive optomechanical coupling via Taylor expansion fails in general. We will return in the following sections to such non-perturbative features, including the topological phase characterizing two-dimensional loops around the DP. On the other hand, for a mechanical mode involving only one of the principal directions (e.g., a sinusoidal mode with $\delta \vec x \propto \hat s$), the regular optomechanical coupling can be characterized by one of the slopes $S_{\pm ,\mu }\equiv \partial k_{\pm }/\partial x_{\mu }$ ($\mu =s,c$). These slopes also determine the anisotropy of the DP spectrum in the branching plane.

For the principal direction $\hat{s}$, Eq.~(\ref{dk_general}) simplifies to: 
\begin{equation}
S_{\pm ,s}=\frac{\sqrt{2N}k_{0}\sin \theta _{m}}{Nd-L_{\rm tot}\left( 1\mp \sqrt{%
1+\sin ^{2}\theta _{m}/\zeta ^{2}}\right) },  \label{sin_mode}
\end{equation}%
which we find also applies to the even/odd modes of a Fabry-Perot cavity (we provide in Methods the derivation of this result for a Fabry-Perot cavity).  We find that the result for $\delta k_+$ in Ref.~\cite{Xuereb2012} differs from Eq.~(\ref{sin_mode}), due to a missing term in the denominator of Eq.~(S.20). However, this discrepancy is not important in the regime considered in Ref.~\cite{Xuereb2012} (see the Methods section for more details). $S_{+,s}$ increases monotonically with $\zeta$ and the $|\zeta|\to +\infty$ limit gives:
\begin{equation}\label{upper_bound}
S_{+ ,s} < 
\frac{k_0}{d}  \sin\left(\frac{m\pi}{N}\right) \sqrt{\frac{2}{N}},
\end{equation}
which is smaller than $k_0/d$ when $N\geq 2$, i.e., the coupling strength cannot
exceed the case of a small cavity constructed by two perfectly reflective
scatterers. Equation~(\ref{upper_bound}) shows that a large value of $N$ is detrimental to the maximum achievable value of $\delta k_+/\delta x_s$. The largest values $\sim k_0/d$ are attained for relatively small $N$, which however also require a large reflectivity ($|\zeta|\gg 1$). 

\begin{figure}
\begin{center}
\includegraphics[width=0.4\textwidth]{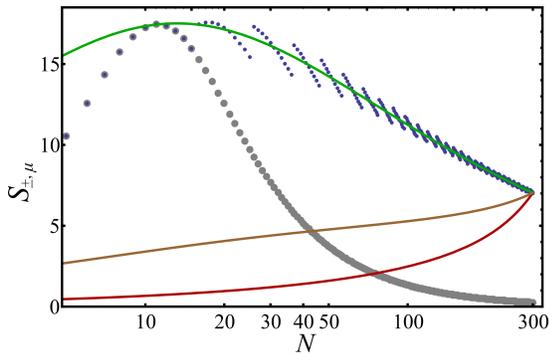}
\end{center}
\caption{Dependence of the enhancement factor $|S_{\pm,\mu}|$ (in units of $k_0/L_{\rm tot}$) 
on the number of scatterers $N$. The three solid lines correspond to 
$S_{+,s}$, $|S_{\pm,c}|$ and $|S_{-,s}|$ (from top to bottom), obtained with $\sin\theta_m=0.3$. 
As $\theta_m=m\pi/N$ is actually a discrete variable, the blue (small) dots show the values of $S_{+,s}$ with $\sin\theta_m$ closest to $0.3$. The dots approaches the green (top) line as $N$ gets larger
and $\sin\theta_m$ becomes continuous. The gray (large) dots are the values of $S_{+,s}$ with $m=1$. 
The other parameters are $\zeta=1$, $L_{\rm tot}/d= 300$. }
\label{opt_stength_fig}
\end{figure}

For scatterers with finite polarizability, the optomechanical coupling is maximized for optimal values of $N,m$, as shown in
Fig.~\ref{opt_stength_fig}. Considering $\zeta \gg \sin \theta _{m}$, the slope $S_{+,s}$ 
first increases with $N$ to reach its maximum value $S_{+,s}^{\max }\simeq k_{0}\zeta /\sqrt{L_{\rm tot}d}$, which agrees very well with
the green (top) solid line in Fig.~\ref{opt_stength_fig}. Hence, as in \cite{Xuereb2012}, the enhancement of $S_{+,s}$ with respect to $%
k_{0}/L_{\rm tot}$ is $\zeta \sqrt{L_{\rm tot}/d}$, i.e., reducing the spacing
between the scatterers helps to increase the optomechanical coupling. Interestingly, we find that the enhancement factor is approximately independent of $m$, although the optimal $N_{\mathrm{opt}}\simeq L_{\rm tot}\sin ^{2}\theta _{m}/(2d\zeta ^{2})$ depends on $m$ explicitly. Therefore, working at $m=1$ is convenient to maximize $S_{+,s}$ with a smaller number of scattering elements.

The slopes $S_{\pm ,c}$ along $ \hat{c}$ can also be found easily from Eq.~(\ref{dk_general}):
\begin{equation}
S_{\pm ,c}=\frac{\pm \sqrt{2N}k_{0}\sin \theta _{m}}{\sqrt{L^2_{\rm tot} \left(
1+\sin ^{2}\theta _{m}/\zeta ^{2}\right)-\left(L_{\rm tot}-Nd\right)^2} },  \label{cos_mode}
\end{equation}%
This expression, together with Eq.~(\ref{sin_mode}), leads to the following inequality (note that $S_{-,s}<0$ and $S_{+,c}=-S_{-,c}$):
\begin{equation}
|S_{-,s}|\leq |S_{\pm ,c}| \leq S_{+ ,s}.  \label{inequality}
\end{equation}
which allows us to apply the upper bound Eq.~(\ref{upper_bound}) also to $|S_{-,s}|, |S_{\pm ,c}|$. We see that the three solid curves of Fig.~\ref{opt_stength_fig} satisfy this inequality and that, at fixed $\theta_m$, both $|S_{-,s}|$ and $|S_{\pm ,c}|$
increase monotonically with $N$ and reach the same value of $S_{+,s}$ when the DBR extends over the whole cavity. 
The case when $d/L_{\rm tot}\ll 1$ and the
DBR is much shorter than the cavity is interesting because the
mode-volume distribution can be substantially modified by the DBR, by becoming more spatially localized. In this regime, the conical intersection is highly anisotropic with
$S_{+,s}\gg |S_{-,s}|, |S_{\pm ,c}|$. This feature is
already evident in Fig.~\ref{opt_stength_fig}. In this limit, the
DP can be approximated as two planes intersecting at the $\delta x_{s}=0$ line, thus the usual form of
optomechanical coupling is recovered (being dominated by $S_{+,s}$). As $S_{+,s}$ coincides with the optomechanical coupling in a Fabry-Perot cavity, this regime is essentially equivalent to the one described in Refs.~\cite{Xuereb2012,Xuereb2013}. This is not surprising, because the boundary condition has a small effect when the light intensity is concentrated within the DBR. We can thus refer to the detailed discussion in Refs.~\cite{Xuereb2012,Xuereb2013} of the effects of absorption and other small imperfections in the DBR, as well as for an analysis of the optical finesse.

Notice finally that, although we have focused on $S_{\pm ,\mu}$, the actual value of the optomechanical couplings
should also take into into account the mechanical mode function $u(x)$ entering the $\delta x_{s,c}$ [since $\delta x_j \propto u(x_j)$]. 
In a ring cavity with sufficiently small indentations \cite{Arbabi2011}, $u(x)$ might be approximately independent of the features of the DBR, but 
the optomechanical coupling stregnth is still strongly dependent on the specific mechanical mode. As the effect of $u(x)$ is strictly related to the implementation details, we considered $S_{\pm ,\mu }= \partial k_{\pm }/\partial x_{\mu }$ in order to characterize the general scale of optomechanical couplings. 

\subsection*{Topological Berry phase} \label{sec:Berry}

Besides the conical energy spectrum, DPs are associated to a topological phase for closed loops around the singularity. This property is not only of considerable theoretical interest, but would allow for a definite experimental identification of the DP. While the degeneracy of the spectrum and the linear dependence of the splitting cannot be observed too close to the DP, due to the finite linewidth, the topological phase is a non-local property which is still valid far from the DP. This is in analogy to the original work of Berry and Wilkinson \cite{Berry1984} where numerical spectra could not allow one to distinguish a degeneracy from a sufficiently small anticrossing. Nevertheless, DPs could be identified  though the presence of a topological phase.

To show this property in our model, we consider the amplitudes $A_{L,R}$ of the eigenvector ${\bf v}_\pm$ of Eq.~(\ref{bc_linearized}). We find that $A^\pm_L/A^\pm_R=e^{i\gamma_\pm }$ where:
\begin{equation}\label{gamma_pm}
e^{i\gamma_\pm }=\left(-1\right) ^{m}
\frac{
\pm\sqrt{\sum_{\mu=s,c} (g_\mu \delta x_\mu)^{2}}-i\frac{\eta\zeta}{\sin \theta _{m}} g_s\delta x_c }
{g_s \delta x_s + \sigma i\left( \frac{\eta\zeta}{\sin\theta _{m}}+\frac{\sin \theta _{m}}{\zeta}\right)\delta x_c}.
\end{equation}
This expression only depends on the polar angle $\phi_{sc}=\arctan{(\delta x_s/\delta x_c)}$. The above form of Eq.~(\ref{gamma_pm}) shows that, if we now consider a loop in the $x_s-x_c$ plane, the numerator has a real part with fixed sign, and always maps the original loop into a trivial loops which does not encircle the origin. On the other hand, the denominator deforms the original loop without changing its topology (i.e., if it encircles the origin or not). The net effect is that a change of $|\Delta\phi_{sc}|=2\pi$ (i.e., a loop around the DP) results in a corresponding change $|\Delta\gamma_\pm|=2\pi$. Since the spatial dependence of the optical field is $\propto \cos \left( k_{\pm }x+\gamma _{\pm}/2\right)$, we recover the well-known $\pi$ phase shift characterizing conical intersections \cite{Berry1984,Berry1984b}. 

For a classical periodic trajectory $\delta\vec{x}(t)$ of the scatterer locations, a phase shift $\pi$ appears for loop B in Fig. \ref{fig:eigenmodes}. Such type of periodic motion could be realized by forced out-of-phase oscillations of two of the scatterers. An even/odd number of periods implies that constructive/destructive interference of the cavity optical field with a reference optical signal could be observed. On the other hand, the topological Berry phase is absent if $\delta\vec{x}(t)$ does not enclose the diabolical point, as for loop A of Fig. \ref{fig:eigenmodes}, and no topological phase is introduced once the mirrors return to the initial configuration in this case. Such topological phase could also be observed through the quantization of the scatterers motion around the origin, as the DB would act similarly as an Aharonov-Bohm flux for charged particles. 

Besides the presence of a topological phase, the DP has significant consequences not only for the adiabatic evolution restricted to the $\pm$ branch, but also for processes involving transitions between the two eigenmodes. This is in analogy to the role played by conical intersections in chemical processes, by allowing non-radiative energy transfer between the molecular energy surfaces. In fact, if we consider a trajectory which crosses $\delta\vec{x}=0$, we see from Eq.~(\ref{gamma_pm}) that a change of sign $\delta x_{s,c}\to -\delta x_{s,c} $ is accompanied by a change of branch to guarantee that the right-hand side does not change discontinuously. Therefore $\gamma_\pm \to \gamma_{\mp}$, which shows that trajectories of type C in Fig. \ref{fig:eigenmodes} can realize an ideal state transfer between the two optical branches. 


\subsection*{Quantum model}\label{sec:quantum}

We would like now to discuss some quantum features of such diabolical points, within a simple optomechanical model. Quantization of the optomechanical interaction is usually achieved by linearization of the frequency dependence on coordinates, $\omega(x)a^\dag a \simeq \omega(x_0)a^\dag a + \omega'(x_0) (x-x_0) a^\dag a$, but this procedure cannot be applied in a straightforward way at a DP. A rigorous quantization of the periodic cavity discussed so far could be accomplished following a similar procedure as Refs.~\cite{Law1994,Law1995}. However, we refrain here from this rather involved task and restrict ourselves to a minimal quantum model, to illuminate the physics. 

As clear from the previous ring resonator analysis, a DP originates from a scattering mechanism between two degenerate optical modes, induced by the motion of at least two of the scatterers. This leads us to consider the following Hamiltonian:
\begin{equation}\label{DP_quantum}
H_{\rm opt}=\Omega \sum_\pm a_\pm^{\dag }a_\pm+\left[(g_{1}x_{1}+g_{2}x_{2})a_+^{\dag }a_-+h.c.\right],
\end{equation}%
where $a_\pm$ are bosonic operators for the clockwise/counterclockwise waves and $x_{1,2}$ are mechanical degrees of freedom (i.e., the positions of two of the $N$ scatterers within the DBR). The interaction term in Eq.~(\ref{DP_quantum}) describes the scattering processes between the clockwise/counterclockwise waves when $x_{1,2} \neq 0$. This is analogous to the linear expansion of the transfer matrix ${\bf M}$ in Eq.~(\ref{dM_linearized}), where a small reflection coefficient is induced by finite $\delta x_i$. The scattering amplitudes $g_{1,2}$ of Eq.~(\ref{DP_quantum}) are in general complex and we require ${\rm Im}[g^*_1g_2]\neq 0$. In this case, the polar angle of $\varphi$ of the total scattering amplitude $g_{1}x_{1}+g_{2}x_{2}$ has a non-trivial dependence on the ratio $x_2/x_1$:
\begin{equation} \label{phi}
\tan\varphi=\frac{\bar{\bar{g}}_1+\bar{\bar{g}}_2 \, x_2/x_1}{\bar{g}_1+\bar{g}_2 \, x_2/x_1},
\end{equation}
where $\bar{g}_{i}=\mathrm{Re}[g_{i}]$ and $\bar{\bar{g}}_{i}=\mathrm{Im}[g_{i}]$.  

For the Hamiltonian in Eq.~(\ref{DP_quantum}), considering for the moment $x_{1,2}$ as classical parameters, the relevant optical eigenmodes $A_\pm$ are easily found as:
\begin{equation}\label{A_pm}
A_\pm =\frac{a_+ \pm a_- e^{i\varphi }}{\sqrt{2}},
\end{equation}%
such that $H_{\rm opt}= \sum_\pm \Omega_\pm A_\pm^{\dag }A_\pm$ with
\begin{equation}\label{Omega_pm}
\Omega _\pm=\Omega \pm \sqrt{(\bar{g}_{1}x_{1}+\bar{g}_{2}x_{2})^{2}+(\bar{\bar{g}}_{1}x_{1}+\bar{\bar{g}}_{2}x_{2})^{2}}.
\end{equation}%
This expression shows that a conical spectrum is obtained in general. If we compare Eq.~(\ref{Omega_pm}) to the linearized spectrum of the ring cavity in Eq.~(\ref{dk_general}), we notice that a term analogous to $(1-\eta)\delta x_s$ is missing here. Such term could be recovered by including linear shifts of the unperturbed frequency in Eq.~(\ref{DP_quantum}), e.g., $\Omega \to \Omega + \sum_i \delta_i x_i$. The coefficients of the quantum model could then be identified with physical parameters in Eq.~(\ref{dk_general}). In the following we restrict ourselves to Eq.~(\ref{DP_quantum}), which already contains the main features of the DP, but our discussion could also be extended to a more complete model. 

The non-analytic feature of the DP is also apparent from Eq.~(\ref{A_pm}). Considering the mechanical motion around an equilibrium position $\vec{x}_0=(x^{(0)}_1,x^{(0)}_2)$, the phase and normal modes in Eq.~(\ref{A_pm}) are well defined ($\varphi \simeq \varphi_0$ and $A_\pm \simeq A^{(0)}_\pm$) only for sufficiently small mechanical displacement around $\vec{x}_0 \neq 0$ (i.e., if $\vec{x}_0$ is away from the DP). On the other hand, when $\vec{x}_0 =0$, $\varphi$ has no definite value no matter how small the mechanical displacement is. Similar considerations hold for the Taylor expansion of Eq.~(\ref{Omega_pm}).

The presence of a topological phase is also easily verified using the Fock states $|n_+,n_- \rangle $ constructed through the $A_\pm^\dag$ operators. The geometrical phase for an adiabatic motion encircling the DP is simply obtained from the integral of the Berry connection:
\begin{equation}\label{Berryphase}
\int_{0}^{2\pi }d\varphi \langle n_+,n_-|\frac{d}{d\varphi }|n_+,n_-\rangle =(n_++n_-)\pi.
\end{equation}
The constant result (independent on the shape of the path) reveals that the Berry curvature is concentrated at the DP, in analogy to an infinitesimally thin solenoid threaded by a half-integer number $(n_++n_-)/2$ of elementary magnetic fluxes. Notice however that the phase factor is not $(-1)$ for an arbitrary quantum state, but depends on the parity of the total number of photons. We can interpret this result by associating the $e^{i\pi}$ phase factor with individual photons, such that the total phase factor is $e^{i\pi(n_++n_-)}$ as in Eq.~(\ref{Berryphase}). The $\pi$ shift of the optical field inside the cavity, which we discussed earlier with classical arguments, can be recovered by considering the transformation of a coherent state $|\alpha \rangle$ of one of the eigenmodes (we suppose the $\pm$ fields to be uncorrelated). It is easily seen that the transformation $|n\rangle \to (-1)^n|n\rangle $ on the Fock states of the given mode corresponds to a $\pi$ phase shift of the coherent state $|\alpha\rangle \to |-\alpha\rangle $. More generally, the effect of the topological phase is to induce the transformation $W(x,p)\to W(-x,-p)$ on the Wigner function of the optical state.

The geometrical phase factor can be understood by introducing the well-known bosonic representation of the angular momentum (Schwinger's
oscillator model):%
\begin{eqnarray}
J_{x} &=&(a_+^{\dag }a_-+a_-^{\dag }a_+)/2, \\
J_{y} &=&-i(a_+^{\dag }a_--a_-^{\dag }a_+)/2, \\
J_{z} &=&(a_+^{\dag }a_+-a_-^{\dag }a_-)/2, \label{Jz}
\end{eqnarray}%
which allows us to write $H_{\rm opt}$ as follows:
\begin{equation}\label{SOI}
H_{\rm opt} = \frac{\Omega J}{2}+2(\bar{g}_{1}x_{1}+\bar{g}_{2}x_{2})J_{x}-2(\bar{\bar{g}}_{1}x_{1}+\bar{\bar{g}}_{2}x_{2})J_{y},
\end{equation}%
with the total angular momentum $J=(n_+ +n_-)/2=0,1/2,1,3/2,\ldots $, a conserved quantity given by the total number of photons.  
In this picture, the displacement of the mechanics away from the DP results in an effective magnetic field coupled to the
effective spin $\vec J$, describing the optical state. For an adiabatic evolution of $x_{1,2}$,
the spin remains aligned with the effective magnetic field and performs a $2\pi$ rotation in the
$x-y$ plane. The geometrical phase of the spin is then $2\pi$ $(\pi)$ for integer (half-integer) spin $J$, i.e., an even (odd) number of photons.

The representation of $H_{\rm opt}$ in Eq.~(\ref{SOI}) is also of interest if the quantization of the mechanical motion is considered, assuming a harmonic unperturbed Hamiltonian $H^{(0)}_{\rm mech}$ with angular frequencies $\omega_{1,2}$ (one can always assume the two modes to have the same mass $m$). By a canonical transformation
\begin{equation}
p_{i} = - (m \omega_i)X_i , \quad x_{i} = (m \omega_i)^{-1} P_i,
\end{equation}
the Hamiltonian $H=H^{(0)}_{\rm mech}+ H_{\rm opt}$ can be written as:
\begin{eqnarray}\label{RD_generalized}
H = && \frac{\Omega J}{2}+\frac{P_{1}^{2}+P_2^2}{2m}+%
\frac{1}{2}m\omega^2_{1}X_1^{2}+\frac{1}{2}m\omega^2_{2}X_2^{2} \nonumber \\
&& +\left( \alpha _{1}P_{1}J_{y}-\alpha _{2}P_{2}J_{x}\right) +\left( \beta
_{1}P_{1}J_{x}-\beta _{2}P_{2}J_{y}\right), \quad
\end{eqnarray}
where we defined the following spin-orbit couplings:
\begin{equation}
\alpha_{i} = -2\bar{\bar{g}}_i(m \omega_i)^{-1}, \quad \beta_{i} = 2\bar{g}_i(m \omega_i)^{-1} .
\end{equation}
Equation (\ref{RD_generalized}) shows that the mechanical motion is equivalent to an anisotropic 2D oscillator in the presence of a spin-orbit interaction [second line of Eq.~(\ref{RD_generalized})] which is reminiscent of a combination of Rashba (when $\alpha _{1}=\alpha_{2}=\alpha $) and Dresselhaus ($\beta _{1}=\beta _{2}=\beta $) couplings \cite{Bychkov1984,Dresselhaus1955}. In fact, by applying independent coordinate and spin rotations (i.e., the rotation angles of $P_{1,2}$ and $J_{x,y}$ will be generally different), the spin-orbit coupling of Eq.~(\ref{RD_generalized}) can be transformed to the canonical form Rashba plus Dresselhaus. For a spin-1/2 particle (the one-photon states of our model), effects of such spin-orbit interactions were studied for a long time in the context of semiconductor quantum dots \cite{Voskoboynikov2001, Aleiner2001}. The Hamiltonian of Eq.~(\ref{RD_generalized}) represents a significant generalization, as any value of the spin is allowed here, depending on the total number of photons. Notice also that the effective crystal axes (determined by the spin-orbit coupling terms) do not coincide in general with the principal axes of the trapping potential, since the coordinate rotation involves the 2D harmonic confinement as well.

For an optomechanical coupling of the form $\sum_{i,j} \eta_{ij} x_i a^\dag_j a_j$, the harmonic excitation spectrum of independent $\hbar\omega_i$ phonons is not affected (the optical cavity only introduces a displaced equilibrium position for oscillator $i$). On the other hand, it can be inferred from numerical studies of quantum dots (see \cite{Marchukov2013} for a recent investigation) that Eq.~(\ref{RD_generalized}) results a rich spectrum for the mechanical system. Similar to Ref.~\onlinecite{Aleiner2001}, we can study the limit of small spin-orbit interactions with the following unitary transformation:
\begin{equation}
U=\exp\left[i \sum_{j=1,2} \frac{\bar{g}_j J_x-\bar{\bar{g}}_j J_y}{\hbar m \omega_j^2/2} p_j \right],
\end{equation}
which eliminates the optomechanical coupling to lowest order. The transformed Hamiltonian $H_{\rm eff}=U^\dag H U$ reads, to second order in $g_i$:
\begin{eqnarray}\label{H_eff}
&& H_{\rm eff} \simeq \frac{\Omega J}{2}+\frac{p_{1}^{2}+p_2^2}{2m} + %
\frac{1}{2}m\omega^2_{1}x_1^{2}+\frac{1}{2}m\omega^2_{2}x_2^{2} \nonumber \\
&&  - \sum_{i} \frac{(\bar{g}_i J_x- \bar{\bar{g}}_i J_y)^2}{m \omega_i^2/2}+{\rm Im}[g^*_1g_2]J_z  \sum_{ij}  \frac{\varepsilon_{ij} x_i p_j}{\hbar m \omega_j^2/2},\qquad
\end{eqnarray}
where $i,j=1,2$ and $\varepsilon_{ij}$ is the antisymmetric tensor with $\varepsilon_{12}=1$. We thus see (in this weak-coupling limit) that the presence of a diabolical point is reflected  in the last term of Eq.~(\ref{H_eff}), which entails an interaction between the two mechanical modes $i,j=1,2$, mediated by the cavity through $J_z$. When ${\rm Im}[g^*_1g_2]=0$, it is not difficult to check using Eqs.~(\ref{phi}-\ref{Omega_pm}) that the more usual optomechanical coupling linear in $x_{1,2}$ is recovered. For example, if for definiteness $g_{1,2}=|g_{1,2}|e^{i\phi}$, the coupling has the form $(\sum_i |g_i|x_i)(A^\dag_+A_+-A^\dag_-A_-)$ with $A_\pm$ independent of $x_{1,2}$ (since $\varphi=\phi$). In this case, the first term in the second line of Eq.~(\ref{H_eff}) is simply the shift in mechanical energy due to the displaced equilibrium position, proportional to $(A^\dag_+A_+-A^\dag_-A_-)^2$, while the second term is absent. 

A particularly transparent result is obtained for two identical oscillators, i.e., $\omega_{1,2}=\omega$. In that case, the last term of Eq.~(\ref{H_eff}) is $\propto J_z L_z$, where the orbital angular momentum $L_z=x_1p_2-x_2 p_1$ is a conserved quantity in the unperturbed Hamiltonian (due to the circular 2D confinement). This type of coupling is well known in quantum dots \cite{Aleiner2001} where the first term in the second line of Eq.~(\ref{H_eff}) is simply a constant, since $J=1/2$. In general the effective Hamiltonian of the optical cavity is given by both terms in the second line of Eq.~(\ref{H_eff}), which only depend on the mechanical angular momentum $L_z$. Conversely, we see that inducing a non-zero value of $\langle J_z \rangle$ favors a certain chirality of the mechanical motion, i.e., a finite value of $L_z$ through the $ \langle J_z \rangle L_z$ coupling. This picture is particularly intuitive in our setup as $\langle J_z \rangle \neq 0$ implies a larger occupation of one of the clockwise/counterclockwise modes of the toroidal cavity [see Eq.~(\ref{Jz})]. This, according to Eq.~(\ref{H_eff}) allows one to transfer angular momentum to the 2D mechanical oscillator.

\section*{Discussion}\label{sec:concl}

We have discussed conical intersections of optical eigenmodes based on a ring cavity with movable scatterers. Characteristic phenomena associated with the presence of DPs, such as the topological Berry phase and adiabatic transduction of photons between optical branches, can be realized in such an optomechanical setup. The Berry phase generated through a DP has a topological character, similar to the Aharonov-Bohm and Aharonov-Casher topological phases, which is distinct from other realizations of geometrical phases in optomechanics \cite{Khosla2013}. Furthermore, Taylor expansion of the cavity frequency on the small mechanical displacements breaks down at the diabolical point due to the non-analytic spectrum and the dependence of the eigenmodes through ratios of the mechanical displacements. Besides a classical analysis of the ring cavity with deformable DBR, we demonstrate the DP features by a minimal quantum model which, interestingly, provides an optomechanical simulation of parabolic quantum dots with spin-orbit interaction. We thus establish a connection with semiconductor physics and quantum gases, where synthetic generation of spin-orbit interactions has recently attracted considerable interest \cite{Galitski2013}. Similarly to the latter case \cite{Lan2014}, the optomechanical model can also extend the well-known semiconductor Hamiltonian to large-spin particles.

A full characterization of the quantized dynamics around the DP seems of particular interest for future investigations. For example, we expect that a more general class of coherent transformations of the optical field can be realized by going beyond the adiabatic limit, since Landau-Zener-type physics becomes relevant slightly away from the degeneracy point.  The creation of entanglement between optical and mechanical degrees of freedom could be realized in analogy to proposals with interacting ultracold atoms \cite{Wuster2011}. However, optomechanical effects  not captured by our Eq.~(\ref{DP_quantum}) could play an important role. Corrections to the non-diagonal terms $a^\dag_+ a_-$ which become larger with the mechanical velocities $d\vec{x}/dt$ were discussed \cite{Law1994,Law1995} and would affect the Landau-Zener transition probability. These effects are usually negligible, due to the large mismatch between optical and mechanical frequencies, but should become relevant in the proximity of the DP. Non-diagonal terms, close to the DP, could also be exploited for photon transduction through a resonant interaction between the mechanical motion and the optical cavity \cite{Law1994}. Finally, the scheme described here is relevant for a variety of optomechanical systems, as in particular micro-toroidal rings with indentations \cite{Arbabi:2011ec} and near-field scatterers \cite{Anetsberger:2009sp,Zhu2010}, but is also applicable to microwave optomechanics ~\cite{Teufel2011} and circuit QED setups (e.g. \cite{Blais2004,Hofheinz2009}).

\section*{Methods}

Making use of the formulas for the ring cavity, we revisit here the problem of a Fabry-Perot cavity with embedded DBR. The transfer matrix $\mathbf{M}(k,\delta \vec{x})$ is the same of the ring cavity, but the boundary condition reads:
\begin{equation}
\mathbf{M}(k,\delta \vec{x}) 
\left(\begin{array}{c}
1 \\
-1
\end{array}\right)=
\left(\begin{array}{c}
a \\
-a
\end{array}\right),
\end{equation}
which, summing the two components and using $M_{11}=M_{22}^*$ and $M_{12}=M_{21}^*$, gives
\begin{equation}
{\rm Im}(M_{11} - M_{12})=0.
\label{bc_fp}
\end{equation}
If we consider that the unperturbed DBR is in the transmissive regime, i.e., $\chi=0$ in Eq.~(\ref{Mt}), we obtain from Eq.~(\ref{bc_fp}) $e^{i\varphi}=\pm 1$, which is related to the parity (even/odd) of the optical eigenmode. Expansion in the small parameters $\delta k,\delta\vec{x}$ gives:
\begin{equation}
\mathbf{M}(k,\delta\vec{x})\simeq \pm {\bf 1}+\sum_{i=1}^N\frac{\partial \mathbf{M}}{%
\partial \delta x_{i}}\delta x_{i}+\frac{\partial \mathbf{M}}{\partial k}\delta
k .
\end{equation}
where $\partial \mathbf{M}/\partial \delta x_{i}$ has the same form of Eq.~(\ref{dMdx}) but the diagonal elements are given by
\begin{equation}
\alpha _{j}e^{ink_{0}L} =\mp 2ik_{0}n\zeta ^{2}\frac{\sin \left(2j-1\right)
\theta _{m}}{\sin \theta _{m}} ,
\end{equation}
while $\beta_j$ is unchanged. Also $\partial \mathbf{M}/\partial k$ has some sign variations from the previous expression  Eq.~(\ref{dMdk}):
\begin{equation}\label{dMdkFP}
\frac{\partial \mathbf{M}}{\partial k}=\left( 
\begin{array}{cc}
\pm i\left( \mu^{\prime }+\chi^{\prime }+nL\right) & i\chi^{\prime } \\ 
-i\chi^{\prime } & \mp i\left( \mu^{\prime }+\chi^{\prime
}+nL\right)%
\end{array}%
\right).
\end{equation}
By using Eq.~(\ref{bc_fp}), we find:
\begin{equation}\label{kpm_fp}
\delta k_\mp =\mp \sum_{j=1}^N \frac{  {\rm Im}(\alpha _{j}e^{ink_{0}L} - \beta_j)}{\mu^{\prime }+\chi^{\prime } \mp \chi^{\prime }+nL}\delta x_j,
\end{equation}
which happen to coincide with Eq.~(\ref{sin_mode}), derived for a periodic cavity. On the other hand, Eq.~(\ref{kpm_fp}) differs from the corresponding expression given in \cite{Xuereb2012}, due to the fact that the $\mu^\prime$ contribution to the denominator of Eq.~(\ref{kpm_fp}) was missed in Eq.~(S.20) of \cite{Xuereb2012}. Nevertheless, this $\mu'$ term has a small effect in certain parameter regimes, considered by \cite{Xuereb2012}. In fact, by taking $\sin\theta^2_m \ll 1,\zeta^2$, we obtain from Eq.~(\ref{sin_mode}) that:
\begin{equation}\label{sin_mode_approx}
\delta k_+ \simeq \frac{2k_0}{L_{\rm tot}}\left(
\frac{\sqrt{2}\zeta^2N^{3/2}/(m\pi)}{1+\frac{2d }{(m\pi)^2 L_{\rm tot}}\zeta^2 N^3} \right) \delta x_s,
\end{equation}
where the only difference of Eq.~(\ref{sin_mode_approx}) from the corresponding expression in \cite{Xuereb2012} is  the presence of $L_{\rm tot}$ instead of $L$, which is not important when $L_{\rm tot}\simeq L$ (i.e., the DBR occupies a negligible portion of the cavity). Since much of the discussion on optomechanical coupling enhancement of \cite{Xuereb2012} refers to this case, it remains mostly valid and can be applied to our periodic cavities as well.

\subsection*{Acknowledgments}

We thank A. A. Clerk for useful discussions. S. C. and Y.-D. W. acknowledge funding from the Youth 1000 Talents of China Program and the Macquarie University Research Centre for Quantum Science and Technology for a Visiting Fellowship. Y.-D. W. acknowledges funding from NSFC (grant No. 11434011). J. T. acknowledges funding from the Australian Research Council Centre of Excellence in Engineered Quantum Systems Project No. CE110001013.

\end{document}